\documentclass{ieee}        	

\usepackage{xspace}
\usepackage{url}
\usepackage{bold-extra}
\usepackage[pdftex,colorlinks=true,pdfstartview=FitV,linkcolor=blue,citecolor= blue,urlcolor= blue]{hyperref}
\usepackage{amsmath}
\usepackage{amssymb}
\usepackage{ifthen}

\usepackage[final]{pdfpages}
\newboolean{showcomments}
\setboolean{showcomments}{false}
\ifthenelse{\boolean{showcomments}}
  {\newcommand{\nb}[2]{
    \fbox{\bfseries\sffamily\scriptsize#1}
    {\sf\small$\blacktriangleright$\textit{#2}$\blacktriangleleft$}
   }
   \newcommand{\version}{\emph{\scriptsize$-$Id$-$}}
  }
  {\newcommand{\nb}[2]{}
   \newcommand{\version}{}
  }

\newcommand{\AK}[1]{\nb{akuhn}{#1}}
\newcommand{\on}[1]{\nb{oscar}{#1}}

\newcommand{\todo}[1]{\nb{todo}{#1}}

\newcommand{\done}[1]{}
\newcommand{\secref}[1]{Section~\ref{sec:#1}}
\newcommand{\figref}[1]{Figure~\ref{fig:#1}}

\newcommand{\MDS}{Multidimensional Scaling\xspace}
\newcommand{\LSI}{Latent Semantic Indexing\xspace}
\newcommand{\TOOL}{\textsc{SoftwareCartographer}\xspace}
\newcommand{\ie}{\emph{i.e.,}\xspace}
\newcommand{\eg}{\emph{e.g.,}\xspace}
\newcommand{\etal}{\emph{et al.}\xspace}

\usepackage[scaled=.92]{helvet}
\usepackage{times}
\usepackage{graphicx}
\usepackage{parskip}
\usepackage[labelfont=bf,textfont=it]{caption}
\title{Consistent Layout for Thematic Software Maps
\thanks{In Proceedings of 15th Working Conference on Reverse Engineering (WCRE'08), IEEE Computer Society Press, Los Alamitos CA, October 2008, pp. 209-218.}\version}
\author{Adrian Kuhn, Peter Loretan, Oscar Nierstrasz \\[0.25cm]
\emph{Software Composition Group, University of Bern, Switzerland}\\
\url{http://scg.unibe.ch}}
\begin{document}
\maketitle
\begin{abstract}
Software visualizations can provide a concise overview of a complex software system.
Unfortunately, since software has no physical shape, there is no ``natural'' mapping of software to a two-dimensional space. As a consequence most visualizations tend to use a layout in which position and distance have no meaning, and consequently layout typical diverges from one visualization to another.
We propose a consistent layout for software maps in which the position of a software artifact reflects its \emph{vocabulary}, and distance corresponds to similarity of vocabulary.
We use \LSI (LSI) to map software artifacts to a vector space, and then use \MDS (MDS) to map this vector space down to two dimensions.
The resulting consistent layout allows us to develop a variety of thematic software maps that express very different aspects of software while making it easy to compare them.
The approach is especially suitable for comparing views of evolving software, since the vocabulary of software artifacts tends to be stable over time.
\end{abstract}


\todo{\\
- checking if the numbers in table 1 are really KLOC (I doubt, but check)\\
- regenerate figure 5 and 6 with labels\\
I just thought that maybe we can do some very simple thematic maps based on the Ludo data\\
--- a view where arrows are used to show inheritance relation between classes\\
--- a view where test tube glyphs are used to indicate test classes and a factory glyph to indicate the factory class, and maybe also some symbol denoting other patterns like a recycle glyph for the iterator class and a puzzle piece glyph the template pattern in the square classes.
}

\section{Introduction}

Software visualization offers an attractive means to abstract from the complexity of large software systems \cite{Dieh07a,Kien07a,Reis05a,Stor05a}.
A single graphic can convey a great deal of information about various aspects of a complex software system, such as its structure, the degree of coupling and cohesion, growth patterns, defect rates, and so on.
Unfortunately, the great wealth of different visualizations that have been developed to abstract away from the complexity of software has led to yet another source of complexity: it is hard to compare different visualizations of the same software system and correlate the information they present.

We can contrast this situation with that of conventional thematic maps found in an atlas.
Different phenomena, ranging from population density to industry sectors, birth rate, or even flow of trade, are all displayed and expressed using \emph{the same consistent layout}.
It easy to correlate different kinds of information concerning the same geographical entities because they are generally presented using the same kind of layout.
This is possible because (i) there is a ``natural'' mapping of position and distance information to a two-dimensional layout (the earth being, luckily, more-or-less flat, at least on a local scale), and (ii) by convention, North is normally considered to be ``up''.\footnote{On traditional Muslim world maps, for example, South used to be on the top. Hence, if Europe would have fallen to the Ottomans at the Battle of Vienna in 1683, all our maps might be drawn upside down \cite{Hite99a}.}

Software artifacts, on the other hand, have no natural layout since they have no physical location.
Distance and orientation also have no obvious meaning for software.
It is presumably for this reason that there are so many different and incomparable ways of visualizing software.
A cursory survey of recent \textsc{Softvis} and \textsc{Vissoft} publications shows that the majority of the presented visualizations feature arbitrary layout, the most common being based on alphabetical ordering and \emph{hash-key ordering}.
(Hash-key ordering is what we get in most programming languages when iterating over the elements of a Set or Dictionary collection.)

Consistent layout for software would make it easier to compare visualizations of different kinds of information, but what should be the basis for laying out and positioning representations of software artifacts within a ``software map''?
What we need is a semantically meaningful notion of position and distance for software artifacts which can then be mapped to consistent layout for 2-D software maps.

We propose to use \emph{vocabulary} as the most natural analogue of physical position for software artifacts, and to map these positions to a two-dimensional space as a way to achieve consistent layout for software maps.
Distance between software artifacts then corresponds to distance in their vocabulary.
Drawing from previous work \cite{Kuhn07a,Duca06c} we apply \LSI to the vocabulary of a system to obtain $n$-dimensional locations, and we use \MDS to then obtain a consistent layout.
Finally we employ digital elevation, hill-shading and contour lines to generate a landscape representing the frequency of topics.

Why should vocabulary be more natural than other properties of source code?
First of all, vocabulary can effectively \emph{abstract} away from the technical details of source code by identifying the key domain concepts reflected by the code \cite{Kuhn07a}.
Software artifacts that have similar vocabulary are therefore close in terms of the domain concepts that they deal with.
Furthermore, it is known that: over time software tends to grow rather than to change \cite{Vasa07b}, and the vocabulary tends to be more stable than the structure of software \cite{Anto07a}. Although refactorings may cause functionality to be renamed or moved, the overall vocabulary tends not to change, except as a side-effect of growth. 
This suggests that vocabulary will be relatively \emph{stable} in the face of change, except where significant growth occurs.
As a consequence, vocabulary not only offers an intuitive notion of position that can be used to provide a consistent layout for different kinds of thematic maps, but it also provides a robust and consistent layout for mapping an evolving system.
System growth can be clearly positioned with respect to old and more stable parts of the same system.

We call our approach \emph{Software Cartography}, and call a series of visualizations \emph{Software Maps}, when they all use the same consistent layout created by our approach.

The contributions of this paper are as follows:
We identify and motivate the need for consistent layouts in software visualization.
We propose a set of techniques to create a consistent layout of a software system: lexical information, LSI, and MDS.
We present \TOOL, a proof-of-concept implementation of software cartography and discuss the algorithms used.
We present examples of thematic software maps that exploit consistent layout to display different information for the same system. 
We also show how consistent layout can be used to illustrate the evolution of a system over time.

The remainder of the paper is structured as follows.
In \secref{techniques} we present our technique for mapping software to consistent layouts.
\secref{casestudy} presents several case studies that illustrate consistent layouts for various thematic software maps.
\secref{related} discusses related work.
Finally, in \secref{conclusion} we conclude with some remarks about future work.

\section{Software Cartography}
\label{sec:techniques}

In this section we present the techniques used to achieve consistent layout for software maps. The number crunching is done by LSI and MSD, whereas the rendering algorithms are mostly from geographic visualization \cite{Sloc05a}. The \TOOL tool\footnote{\url{http://smallwiki.unibe.ch/softwarecartography/}} provides a proof-of-concept implementation of our technique. 

\subsection{Lexical similarity}

In order to define a consistent layout for software visualization, we need the position of software artifacts and the distance between them to reflect a natural notion of position and distance in reality.

Instead of looking for distance metrics in the graph structure of programs, we propose to focus on the vocabulary of source code artifacts as the space within which to define their position and distance.
Lexical similarity denotes how close software artifacts are in terms of their source code's vocabulary.
The vocabulary of the source code corresponds to the implemented technical or domain concepts.
Artifacts with similar vocabulary are thus conceptually and topically close \cite{Kuhn07a}.

\LSI (LSI) is an information retrieval technique originally developed for use in search engines, with applications to software analysis \cite{Marc08a}.
Since source code consists essentially of text, we can apply LSI to source code to retrieve the lexical distance between software artifacts.

The examples in this paper use the \textsc{Hapax} tool\footnote{\url{http://smallwiki.unibe.ch/adriankuhn/hapax/}} to compute the lexical similarity between software artifacts.
Given a software system $\mathcal{S}$ which is a set of software entities $s_1 \dots s_n$ using terms $t_1 \dots t_m$, then \textsc{Hapax} uses LSI to generate an $m$-dimensional vector space $\mathbb{V}$ representing the lexical data of the software system.

In this vector space $\mathbb{V}$, each software entity is represented by a vector of its term frequencies. Thus, in information retrieval, $\mathbb{V}$ is often referred to as ``term-document matrix''.

The terms $t_1 \dots t_m$ are the identifiers found in the source code: class names, methods names, parameter names, local variables names, names of invoked methods, et cetera. Thus, two documents (\ie source files or classes) are not only similar if they are structurally related, but also if they use the same identifiers only. This has proven useful to detect high-level clones\cite{Marc01a} and cross-cutting concerns\cite{Kuhn07a}.

\subsection{Multidimensional scaling}

The elements shown on the software map are the software entities $s_1 \dots s_n$ labeled with their file or class name. The visualization pane is two-dimensional, whereas LSI locates all software entities in an $m$-dimensional space. Hence, we must map positions in $\mathbb{V}$ down to two dimensions.
There are three main techniques to do this, each of which is suitable for very different purposes:

\begin{enumerate}
  \item \emph{Principal Component Analysis} (PCA) is perhaps the most widely used of all three techniques. PCA yields the best low-level approximation with regard to variance and classification. It tries to preserve as much of the space's variance in the remaining dimensions. Hence, PCA is the best choice for classification problems.

  \item \emph{Singular-Value Decomposition} (SVD) treats the $m$-dimensional space as matrix $A_{n \times m}$, which is the mathematical equivalent of an $m$-dimensional vector space with $n$ vectors. SVD yields the best low-rank approximation $A'$ under a least-squares criterion. It tries to preserve as much of the space's eigenvalue. Hence, SVD is the best choice for lossy compression and signal reduction problems. 

  \item \emph{\MDS} (MDS) tries to minimize a stress function while iteratively placing elements into a low-level space. MDS yields the best approximation of a vector space's orientation, \ie preserves the relation between elements as best as possible. Hence, MDS is the best choice for data exploration problems.

\end{enumerate}

For the purpose of software cartography, preserving the relative lexical similarity of software entities is most important. Thus, MDS is the best choice for mapping the LSI vector space to our target visualization space.

MDS attempts to arrange objects in a low-dimensional space so that the distance between them in the target space reflects their similarity. The input for the algorithm is an $n \times n$ similarity square matrix, where $n$ is equal to the number of objects to display. Each cell $(x, y)$ of the matrix contains the similarity between object~$x$ and object~$y$. The dimension of the solution space can range from $2$ to $n-1$.
As the number of target dimensions decreases, clearly the quality of the approximation deteriorates.

In our case we feed MDS with the lexical similarity of software artifact, and map them on a $2$-dimensional visualization space. When computing the similarity of lexical data, it is important to use a cosine or Pearson distance metric, as the standard Euclidian distance has no meaningful interpretation when applied to documents and term-frequencies!

\newlength{\figwidth}


\begin{figure}
\begin{center}
  \fbox{\includegraphics[width=0.9\linewidth]{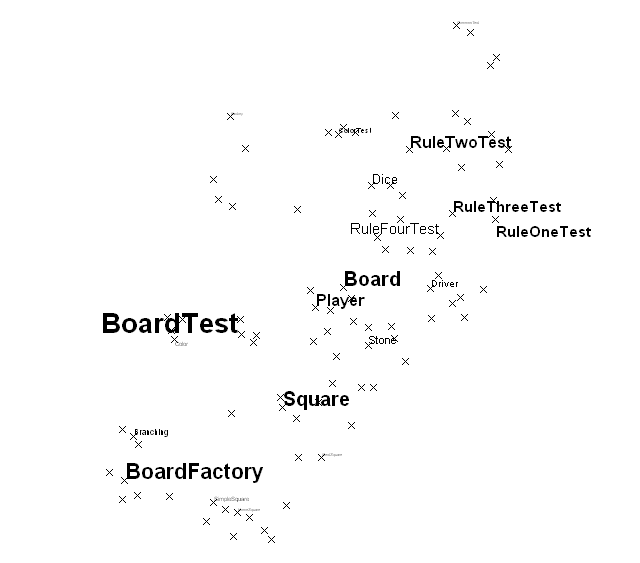}}\\
  \fbox{\includegraphics[width=0.9\linewidth]{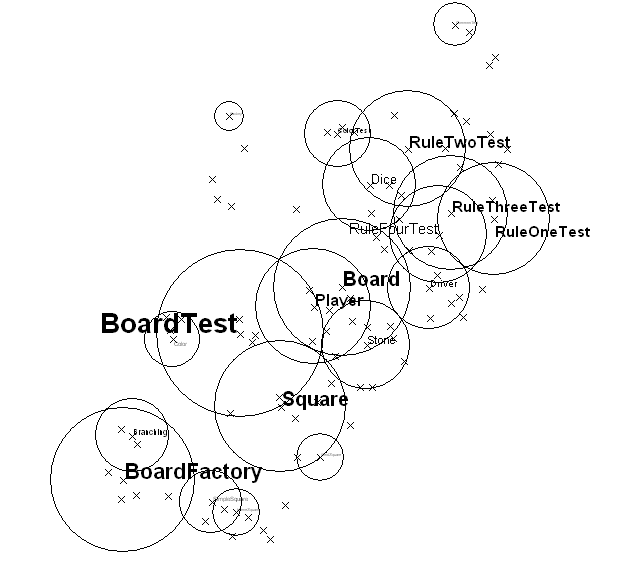}}\\
  \fbox{\includegraphics[width=0.9\linewidth]{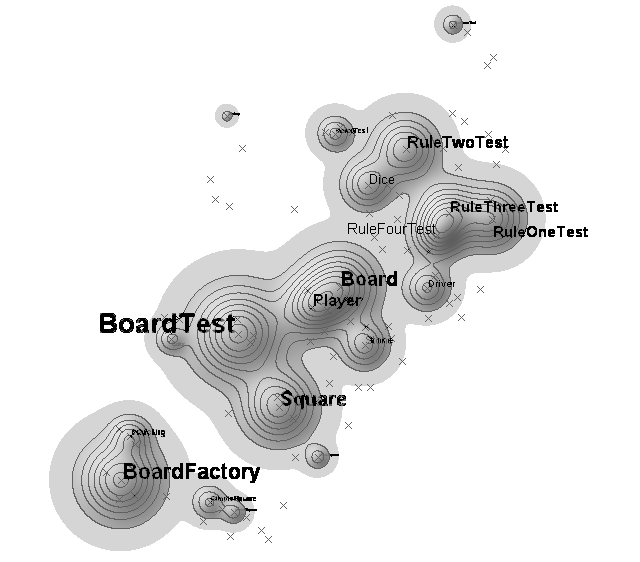}}
\end{center}
    \caption{Construction steps of a software map. From top to bottom: 1) dots in the visualization space, positioned with MDS, 2) circles around each entity's location, based on class size in KLOC, 3) digital elevation model with hill-shading and contour lines.
    }
    \label{fig:steps}
\end{figure}

\subsection{Iterative scaling}

MDS is an iterative algorithm. Given an as an input the similarity between objects, it works as follows:
\begin{enumerate}
\item Assign all objects an arbitrary location in the solution space.
\item Determinate the goodness of fit, \ie compare the distance between the objects in the solution space with their similarities given in the input.
\item If the stress value, \ie the goodness of fit, is within a given threshold, terminate.
\item Search for a monotonic transformation of the data. That is, far apart but similar objects are moved towards each other, close but not similar objects are moved away from each other. Proceed with step 3.
\end{enumerate}

\AK{Peter: is the above correct?}

During the second step it is important to have a good measure  for the goodness of the approximation. For this reason the stress value was introduced, which shows the natural goodness of a configuration as a single number. A stress value of 0 stands for an optimal solution where the distances between the objects in the configuration perfectly fits their dissimilarity. A higher stress value indicates an increased approximation level between distances and dissimilarities. 



In the \TOOL tool, we apply High-Throughput MDS (HiT-MDS), which is an optimized implementation of MDS particularly suited for dealing with large data sets \cite{Stri05a}. The algorithm was originally designed for clustering multi-parallel gene expression probes. These data sets contain thousands of gene probes and the corresponding similarity matrix dimension reflects this huge data amount. The price paid for a fast computation is less accurate approximation and a simplified distance metric.

As a consequence of these optimizations, the generated output may vary when run several times on the same input, \ie  HiT-MDS uses non-deterministic heuristics. In practice, this appears to be good enough for our experiments with \TOOL and software analysis. 

\subsection{Hill-shading and Contour Lines}
In \figref{steps} we see an overview of the steps taken to render a software map.
To make our map more aesthetically appealing, we add a touch of three-dimensionality.

The hill-shading algorithm is well-known in geographic visualization. It adds hill shades to a map \cite{Sloc05a}. The algorithm works on a distinct height model (digital elevation model) rather than on trigonometric data vetor date: each pixel has an assigned z-value, its height. 

\begin{figure}[t]
\centering
\includegraphics[width=\linewidth]{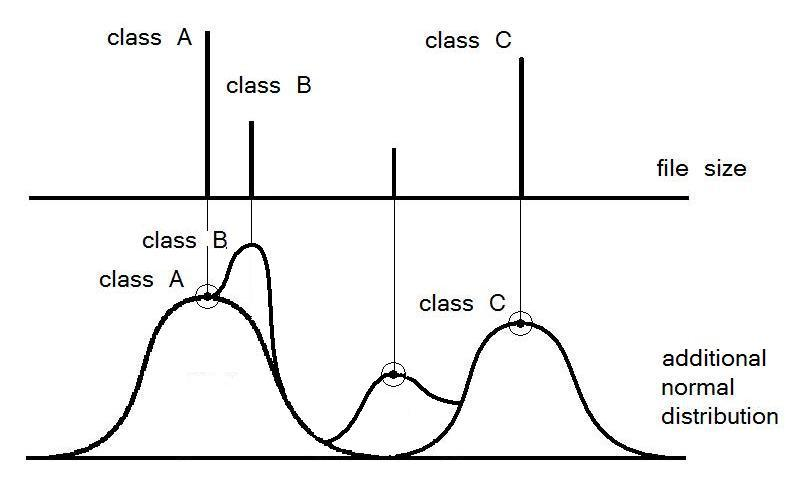}
\caption{Digital elevation model: each element is represented by a normal distribution according to its KLOC size, the distribution of all elements is summed up.
\on{The resolution of the figure is extremely poor! Where is the original figure?} \AK{Peter, can you help?}
}
\label{fig:dem}
\end{figure}

The digital elevation model of \TOOL is is a simple matrix with discrete height information for all pixels of the visualization plane. As illustrated on Figure~\ref{fig:dem}, each element (ie source file of class) is represented by the a hill who's height corresponds to the element's KLOC size. The shape of the hill is determined using a normal distribution function. To avoid that closely located element hide each other, the elevation of all individual elements is summed up.

The hill-shading algorithm renders a three-dimensional looking surface by determining an illumination value for each cell in that matrix. It does this by assuming a hypothetical light source and calculating the illumination value for each cell in relation to its neighboring cells.

Eventually, we add contour lines. Drawing contour lines on maps is a very common technique in cartography. Contour lines make elevation more evident then hill-shading alone. Since almost all real world maps make use contour lines, maps with contour lines are very familiar to the user.

\begin{figure*}[t]
  \includegraphics[width=\linewidth]{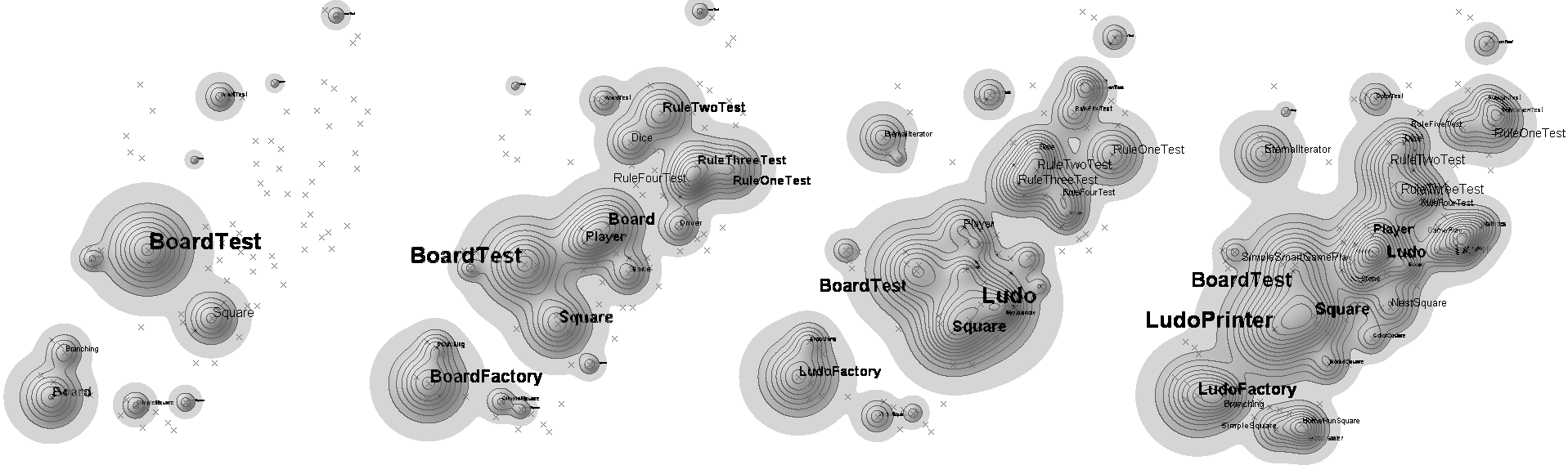}
  \caption{From left to right: each map shows an consecutieve iteration of the same software system. As all four views use the same layout, a user can build up a mental model of the system's spatial structure. For example, {\tt Board/LudoFactory} is on all four views located in the south-western quadrant. See also Figure~5 and 6 for more views of this system.}
  \label{fig:ludo}
\end{figure*}

\subsection{Labeling}

A map without labels is of little use. On a software map, all entities are labeled with their name (class or file name).

Labeling is a non-trivial problem, we must make sure that no two labels overlap. Also labels should not overlap important landmarks. Most labeling approaching are semi-automatic and need manual adjustment, an optimal labeling algorithm does not exist \cite{Sloc05a}. For locations that are near to each other it is difficult to place the labels so that they do not overlap and hide each other. For software maps it is even harder due to often long class names and clusters of closely related classes.

The examples given in this paper show only the most important class names. \TOOL uses fully-automatic, greedy brute-force approach. Labels are placed either to the top left, top right, bottom left, or bottom right of their element. Smallers labels are omitted if covered by a larger label. Eventually, among all layouts, the one where most labels are shown is chosen. 
\section{Case study}
\label{sec:casestudy}

This section presents examples of software maps. Our first example visualizes the evolution of a software system to illustrates the consisten layout of software maps. Second, we show an overview of six open-source systems to illustrate their distinct spatial layouts. And we present three examples of thematic cartography.

\subsection{Ludo example}

Figure~\ref{fig:ludo} shows the complete history of the Ludo system, consisting of four iterations. Ludo is used in a first year programming course to teach iterative development. The 4th iteration is the largest with 30 classes and a total size of 3-4 KLOC. We selected Ludo because in each iteration, a crucial part of the final system is added. 

\begin{itemize}

\item The first map (Figure~1, leftmost) shows the initial prototype. This iteration implements the board as a linked list of squares. Most classes are located in the south-western quadrant. The remaining space is occupied by ocean, nothing else has been implemented so far.

\item In the second iteration (Figure~3, second to the left) the board class is extended with a factory class. In order to support players and stones, a few new classes and tests for future game rules are added. On the new map the test classes are positioned in the north-eastern quadrant, opposite to the other classes. This indicates that the newly added test classes implement a novel feature (\ie testing of the game's ``business rules'') and are thus not related to the factory's domain of board initialization. 

\item During the third iteration (Figure~3, second to the right) the actual game rules are implemented. Most rules are implemented in the {\tt Square} and {\tt Ludo} class, thus their mountain rises. In the south-west, we can notice that, although the {\tt BoardFactory} has been renamed to {\tt LudoFactory}, its position on the map has not changed considerably. 

\item The fourth map (Figure~3, rightmost) shows the last iteration. A user interface and a printer class have been added. Since both of them depend on most previous parts of the application they are located in the middle of the map. As the UI uses the vocabulary of all different parts of the system, the islands start to grow together.

\end{itemize}

The layout of elements remains stable over all four iterations.  For example, {\tt Board/LudoFactory} is on all four views located in the south-western quadrant. This is due to LSI's robustness in the face of synonymy and polysemy; as a consequence most renamings do not significantly change the vocabulary of a software artifact \cite{Kuhn07a}.

\setlength{\figwidth}{0.32\textwidth}
\begin{figure*}
\begin{center}
\begin{minipage}{\figwidth}
\begin{center}
  \includegraphics[width=\figwidth]{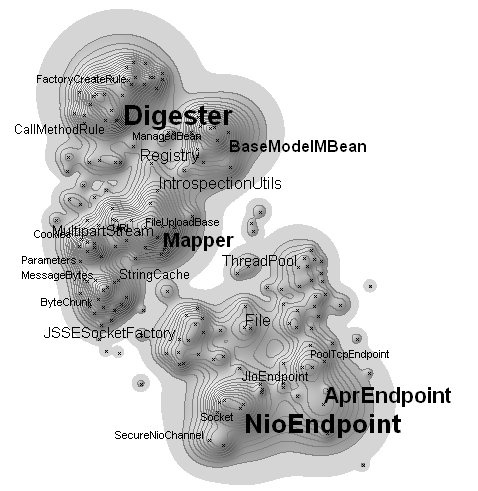}\\
  Apache Tomcat
\end{center}
\end{minipage}~
\begin{minipage}{\figwidth}
\begin{center}
  \includegraphics[width=\figwidth]{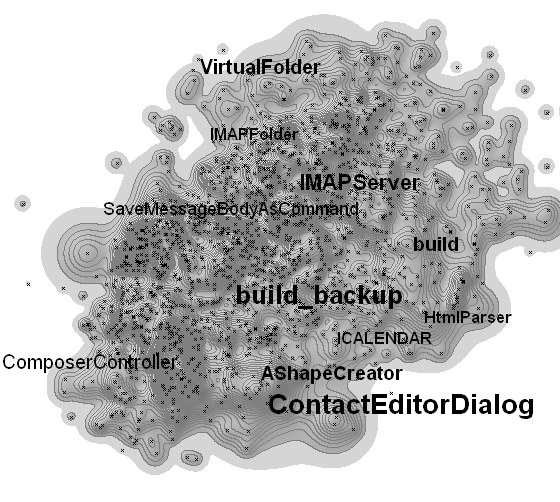}\\
  Columba
\end{center}
\end{minipage}
\begin{minipage}{\figwidth}
\begin{center}
  \includegraphics[width=\figwidth]{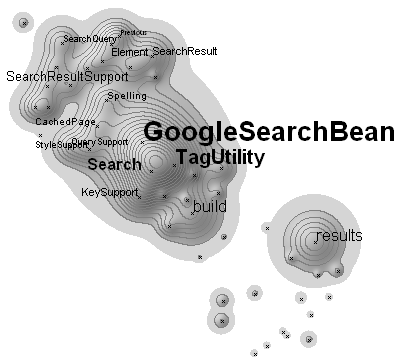}\\
  Google Taglib
\end{center}
\end{minipage}
\begin{minipage}{\figwidth}
\begin{center}
  \includegraphics[width=\figwidth]{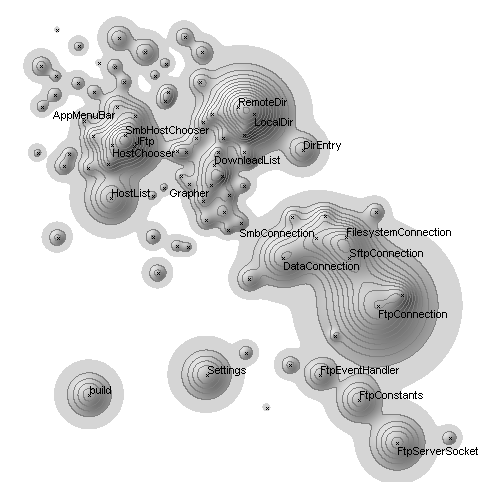}\\
  JFtp
\end{center}
\end{minipage}
\begin{minipage}{\figwidth}
\begin{center}
  \includegraphics[width=\figwidth]{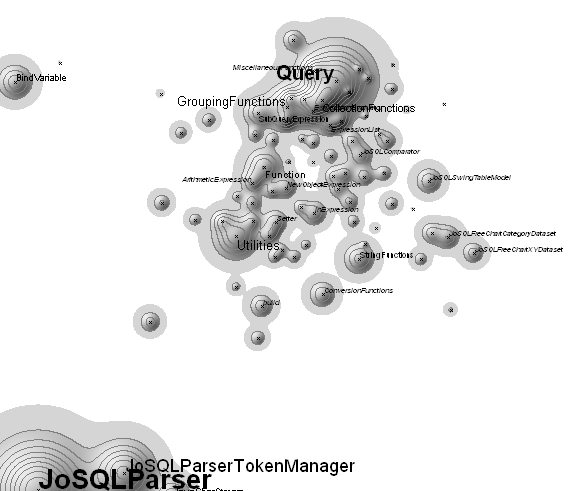}\\
  JoSQL
\end{center}
\end{minipage}
\begin{minipage}{\figwidth}
\begin{center}
  \includegraphics[width=\figwidth]{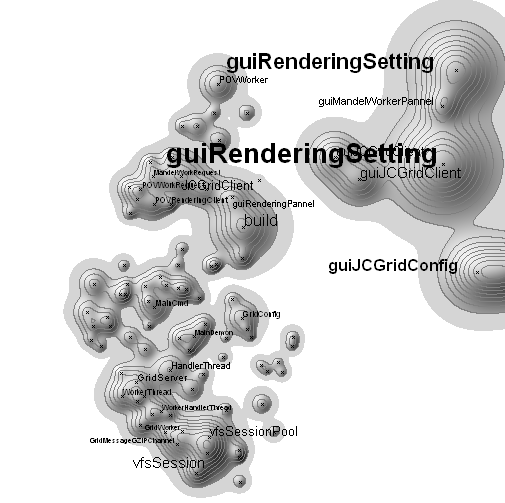}\\
  JCGrid
\end{center}
\end{minipage}
\end{center}
\vspace{1ex}
    \caption{Overview of the software maps of six open source systems. Each map reveals a distinct spatial structure. When consequently applied to every visualization, the consistent layout may soon turn into the system's iconcic fingerprint. An engineer might \eg point to the top left map and say: ``Look, this huge {\tt Digester} peninsula in the north, that must be Tomcat. I know it from last year's code review.''.
    }
    \label{fig:fullpage}
\end{figure*} 

\subsection{Open-source examples}

We applied the software cartography approach to all systems listed in the field study by Cabral and Marques \cite{Cabr07a}. They list 32 systems, including 4 of each type of application (Standalone, Server, Server Applications, Libraries) and selected programming language (Java, .NET).  

Figure \ref{fig:fullpage} shows the software map for six of these systems: Apache Tomcat, Columba, Google Taglib, JFtp, JCGrid and JoSQL. Each system reveals a distinct spatial structure. Some fall apart into many islands, like JFtp, whereas others cluster into one (or possibly two) large contents, like Columba and Apache Tomcat. The 36 case-studies raised interesting questions for future work regarding the correlation between a system's layout and code quality. For example, do large continents indicate bad modularizations? Or, do archipelagoes indicate low coupling?



\subsection{Thematic cartography examples}

\begin{figure}[t]
\begin{center}
  \includegraphics[width=\linewidth]{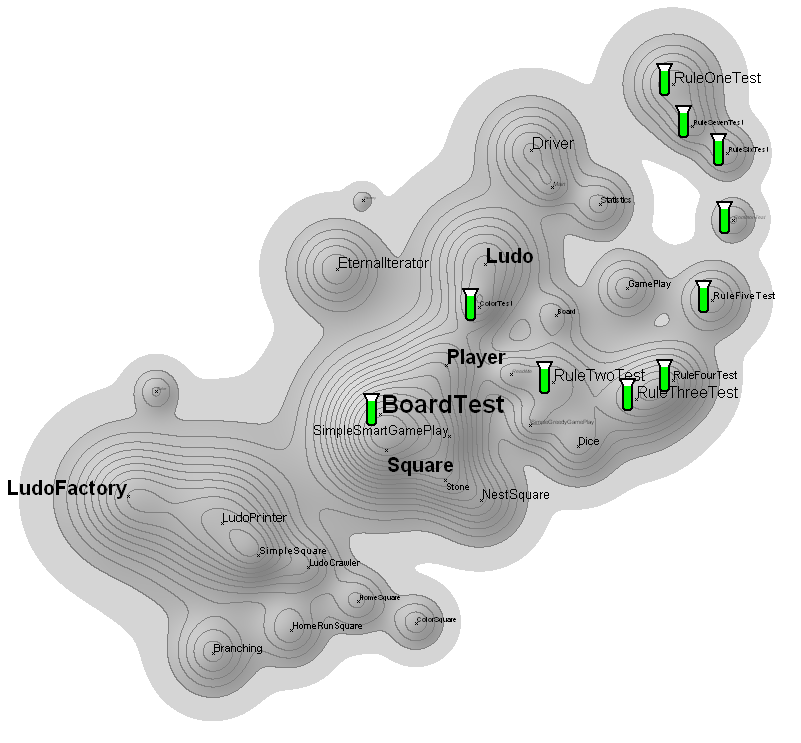}\\[1ex]
\end{center}
\vspace{1ex}
    \caption{Glyphs are drawn on top of the map, to display additional information. Each test tube glyph indicates the location of unit test case.}
    \label{fig:mock-glyphs}
\end{figure}

\begin{figure}[t]
\begin{center}
  \includegraphics[width=\linewidth]{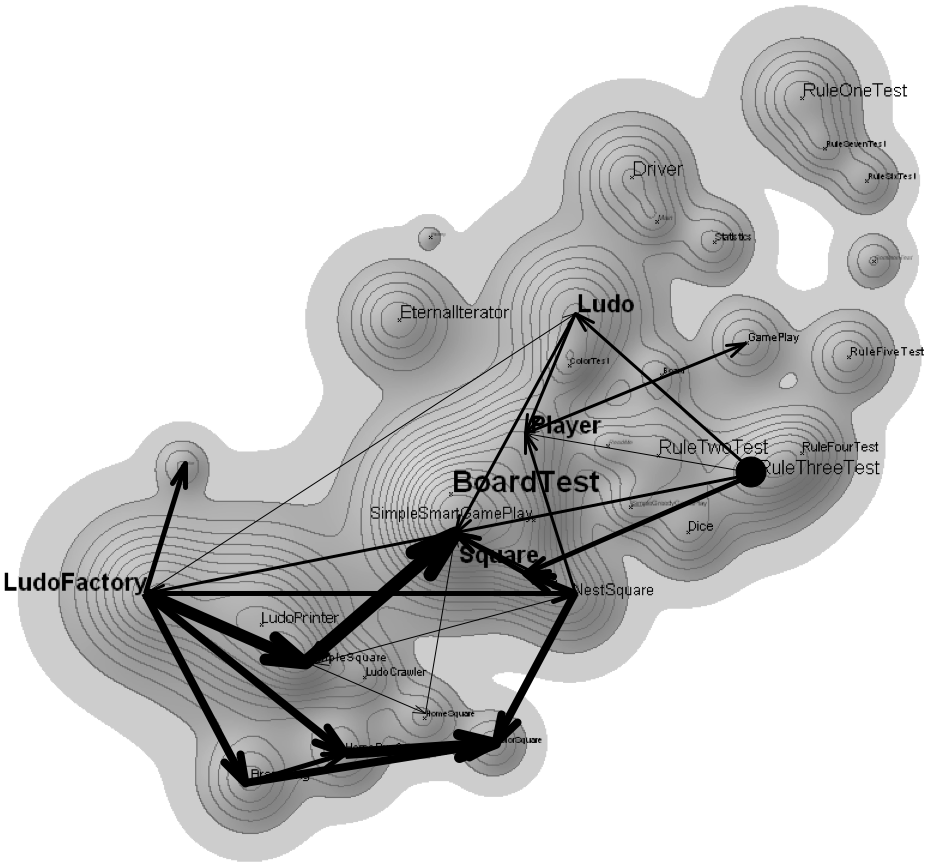}
\end{center}
\vspace{1ex}
    \caption{Invocation edges are drawn on top of the map, showing the trace of executing the {\tt RuleThreeTest} test case.
    }
    \label{fig:mock-invocation}
\end{figure}

Software maps can be used as canvas for more specialized visualizations of the same system. In the following, we provide two thematic visualization of the Ludo system that might benefit from consistent layout. (The maps in this subsection are mockups, not yet fully supported by \TOOL.)

\begin{itemize}

\item Boccuzzo and Gall present a set of metaphors for the visual shape of entities \cite{Bocc07a}. They use simple and well-known graphical elements from daily life, such as houses and tables. However they use conventional albeit arbitrary layouts, where the distribution of glyphs often does not bear a meaningful interpretation. The first map in Figure~\ref{fig:mock-glyphs} employs their technique on top of a software map, using test tubes to indicate the distribution of test cases.

\item Greevy \etal present a three-dimensional variation of System Complexity View to visualize a System's dynamic runtime state \cite{Gree05d}. They connect classes with edges representing method invocation, and stack boxes on top of each other to represent a class's instances.
Since System Complexity Views do not capture any notion of position, the lengths of their invocation edges do not express any real sense of distance.

Figure~\ref{fig:mock-invocation} employs their approach on top of a software map, drawing invocation edges in a two-dimensional plane.
Here the distances have an interpretation in terms of lexical distance, so the lengths of invocation edges are meaningful.
A short edge indicates that closely related artifacts are invoking each other, whereas long edges indicate a ``long-distance call'' to a lexically unrelated class.

\end{itemize}

\begin{figure}[t]
\begin{center}
  \includegraphics[width=\linewidth]{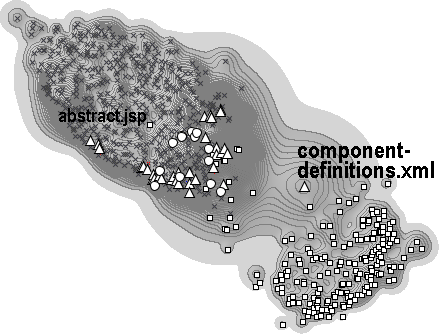}
\end{center}
\vspace{1ex}
    \caption{The KMUadmin JSP application.
    Java files are displayed as crosses, JSP files as squares, XML files as triangles, and property files as circles.
  \AK{Please add labels. How shall a reader interpret the meaning of eg north-east without labels?}
    }
    \label{fig:kmuadmin}
\end{figure}

In \figref{kmuadmin} we see an industrial J2EE application in which artifacts are colored according to which kinds of files they are.
Java source files are crosses, and are all in the north west region.
JSP files are squares and mostly reside in the south east corner of the island.
XML files are triangles and property files are circles.
Both of these are mostly in the central region.

Since Java files have to do with the underlying implementation, and JSP files are closer to domain concepts, it is not too surprising that these files are mostly in separate parts of the island.
Strangely, however, we do find a number of JSP files in the north west, which could mean that they  are more closely linked to the implementation.
One of these (\verb|abstract.jsp|), is interesting because it contains a large portion of Java code and only a small amount of html/jsp.
This explains why its vocabulary places it squarely in the Java part of the island.

Also unusual is a single, isolated XML file on a mountain peak in the south west.
Again, the name (\verb|component-definitions.xml|) suggest that it acts as a bridge between the two regions of the island. In fact, it references many of the JSP files, but is also quite large, and it contains many XML tags and attributes which occur nowhere else, which explains why it stands alone on top of a large hill.

\section{Related work}
\label{sec:related}

Using MDS to create a map of information is by no means a novel idea.
Topic maps, as they are called, have a long standing tradition in information visualization \cite{Ware04a}. 

ThemeScape is the best-known example of a text visualization tool that organizes topics found in documents into topic maps where physical distance correlates to topical distance and surface height corresponds to topical frequency \cite{Wise99a}.
ThemeScape is part of a larger toolset that uses a variety of algorithms to cluster terms in documents.
It then uses MDS for smaller document sets or their own proprietary algorithm, called ``Anchored Least Stress'', for larger document sets to project vector spaces to two dimensions.
The landscape is then constructed by successively layering the contributions of the contributing topical terms, similar to our approach \cite{Wise99a}.

Topic maps in general and Themescape-style maps are rarely used in the software visualization community.
We are unaware of their application in software visualization to produce consistent layouts for thematic maps, or to visualize the evolution of a software system.

Most software visualization layouts are based on one or multiple of the three following approaches: 1) UML diagrams, 2) force-based graph drawing, and 3) tree-map layouts.

\paragraph{\bf UML diagrams.}
UML diagrams generally employ arbitrary layout.
Gudenberg \etal have proposed an evolutionary approach to layout UML diagrams in which a fitness function is used to optimize various metrics (such as number of edge crossings) \cite{Gude06a}. 
Although the resulting layout does not reflect a distance metric, in principle the technique could be adapted to do so.
Achieving a consistent layout is not a goal in this work.

Andriyevksa \etal have conducted user studies to assess the effect that different UML layout schemes have on software comprehension \cite{Andr05a}.
They report that the layout scheme that groups architecturally related classes together yields best results.
They conclude that it is more important that a layout scheme convey
a meaningful grouping of entities, rather than being aesthetically appealing.

Byelas and Telea highlight related elements in a UML diagram using a custom ``area of interest'' algorithm that connects all related elements with a blob of the same color, taking special care to minimize the number of crossings \cite{Byel06a}.
The impact of an arbitrary layout on their approach is not discussed.

\paragraph{\bf Graph drawing.}
Graph drawing refers to a number of techniques to layout two- and three-dimensional graphs for the purpose of information visualization \cite{Ware04a,Kauf01b}. Noack \etal offer a good starting point for applying graph drawing to software visualization \cite{Noac05a}.

Unlike MDS, graph drawing does not attempt to map an $n$-dimensional space to two dimensions, but rather optimizes a fitness function related to the spatial property of the output, \ie of the visualization. Force-based layout for example, tries to minimize the number of edge crossings and to place all nodes as equally apart from each other as possible. 

Jucknath-John \etal present a technique to achieve stable graph layouts over the evolution of the displayed software system \cite{Juck06a}, thus achieving consistent layout, while sidestepping the issue of reflecting meaningful position or distance metrics.

\paragraph{\bf Treemap layout}

Treemaps represent tree-structured information using nested rectangles \cite{Ware04a}.
Though treemaps can achieve a consistent layout, position and distance are not meaningful.
First of all, they are often applied with arbitrary order of elements within packages, \ie alphabetical order. Second, the layout algorithm does not guarantee any spatial constraints between the leaf packages contained in packages that touch at a higher level.
Treemaps may contain very narrow and distorted rectangles.

Balzer \etal proposed a modification of the classical treemap layout using Voronoi tessellation \cite{Balz05a}. Their approach creates aesthetically more appealing treemaps, reducing the number of narrow tessels.

\paragraph{\bf Cartography metaphors for software}

A number of tools have adopted metaphors from cartography in recent years to visualize software.
Usually these approaches are integrated in a tool with in an interactive, explorative interface and often feature three-dimensional visualizations.

MetricView is an exploratory environment featuring UML diagram visualizations \cite{Term05a}. The third dimension is used to extend UML with polymetric views \cite{Lanz03d}.
The diagrams use arbitrary layout, so do not reflect meaningful distance or position.

White Coats is an explorative environment also based on the notion of polymetric views \cite{Mesn05b}. The visualizations are three-dimensional with position and visual-distance of entities given by selected metrics. However they do not incorporate the notion of a consistent layout.

CGA Call Graph Analyser is an explorative environment that visualizes a combination of function call graph and nested modules structure \cite{Bohn07a}. The tool employs a 2$\frac{1}{2}$-dimensional approach. To our best knowledge, their visualizations use an arbitrary layout.

CodeCity is an explorative environment building on the city metaphor \cite{Wett07b}. CodeCity employs the nesting level of packages for their city's elevation model, and uses a modified tree layout to position the entities, \ie packages and classes. Within a package, elements are ordered by size of the element's visual representation. Hence, when changing the metrics mapped on width and height, the overall layout of the city changes, and thus, the consistent layout breaks.

VERSO is an explorative environment that is also based on the city metaphor \cite{Lang05a}. Similar to CodeCity, VERSO employs a treemap layout to position their elements. Within a package, elements are either ordered by their color or by first appearance in the system's history. As the leaf elements have all the same base size, changing this setting does not change the overall layout. Hence, they provide consistent layout, however within the spatial limitations of the classical treemap layout.

\section{Conclusion}
\label{sec:conclusion}

We have presented an approach to visualizing software based on a cartography metaphor, in which \LSI is used to position the \emph{vocabulary} of software entities in an $m$-dimension space, and \MDS is then used to map these positions to a two-dimensional display.
Digital elevation, hill-shading and contour lines are applied to produce a software map.
Finally, software maps can be generated to depict evolution over time of a software system, or they may be decorated to present various kinds of additional thematic information, such as package structure or call relationships.

In spite of the aesthetic appeal of hill shading and contour lines, the main contribution of this paper is not that the visualizations looks like a cartographic map, but rather that (i) cartographic position and distance reflect topical position and distance for software entities, and (ii) consistent layout allows different software maps to be easily compared.
In this way, software maps reflect world maps in an atlas that exploit the same consistent layout to depict various kinds of thematic information about geographical sites.

We have presented several examples to illustrate the usefulness of software maps to depict the evolution of software systems, and to serve as a background for thematic visualizations.
The examples have been produced using \TOOL, a proof-of-concept tool that implements our technique.

As future work, we can identify the following promising directions:
\begin{itemize}
  \item Software maps at present are largely static.
  We envision a more interactive environment in which the user can ``zoom and pan'' through the landscape to see features in closer detail, or navigate to other views of the software.
  \item Selectively displaying features would make the environment more attractive for navigation. Instead of generating all the labels and thematic widgets up-front, users can annotate the map, adding comments and waymarks as they perform their tasks.
  \item Orientation and layout are presently consistent for a single project only.
  We would like to investigate the usefulness of conventions for establishing consistent layout and orientation (\ie ``testing'' is North-East) that will work across multiple projects, possibly within a reasonably well-defined domain.
  \item We plan to perform an empirical user study to evaluate the application of software cartography for software comprehension and reverse engineering, but also for source code navigation in development environments.
\end{itemize}

\section*{Acknowledgements}

We thank Elias Hodel for his help with the hill-shading and contour line algorithms, and Toon Verwaest for his constructive review.
We also thank Tudor G\^irba for his comments and his help in preparing the J2EE case study.

We gratefully acknowledge the financial support of
the Hasler Foundation for the project ``Enabling the evolution of J2EE applications through reverse engineering and quality assurance''
and
the Swiss National Science Foundation for the project ``Analyzing, Capturing and Taming Software Change" (SNF Project No.\ 200020-113342, Oct.\ 2006 - Sept.\ 2008).

\bibliographystyle{plain}
\bibliography{scg}

\begin{thebibliography}{10}

\bibitem{Andr05a}
Olena Andriyevska, Natalia Dragan, Bonita Simoes, and Jonathan~I. Maletic.
\newblock Evaluating {UML} class diagram layout based on architectural
  importance.
\newblock {\em VISSOFT 2005. 3rd IEEE International Workshop on Visualizing
  Software for Understanding and Analysis}, 0:9, 2005.

\bibitem{Anto07a}
Giuliano Antoniol, Yann-Gael~Gueh eneuc, Ettore Merlo, and Paolo Tonella.
\newblock Mining the lexicon used by programmers during sofware evolution.
\newblock In {\em ICSM 2007: IEEE International Conference on Software
  Maintenance}, pages 14--23, October 2007.

\bibitem{Balz05a}
Michael Balzer, Oliver Deussen, and Claus Lewerentz.
\newblock Voronoi treemaps for the visualization of software metrics.
\newblock In {\em SoftVis '05: Proceedings of the 2005 ACM symposium on
  Software visualization}, pages 165--172, New York, NY, USA, 2005. ACM.

\bibitem{Bocc07a}
Sandro Boccuzzo and Harald Gall.
\newblock {CocoViz}: Towards cognitive software visualizations.
\newblock {\em VISSOFT 2007. 4th IEEE International Workshop on Visualizing
  Software for Understanding and Analysis}, 0:72--79, 2007.

\bibitem{Bohn07a}
Johannes Bohnet and Jurgen Dollner.
\newblock {CGA} call graph analyzer --- locating and understanding
  functionality within the {Gnu} compiler collection's million lines of code.
\newblock {\em VISSOFT 2007. 4th IEEE International Workshop on Visualizing
  Software for Understanding and Analysis}, 0:161--162, 2007.

\bibitem{Byel06a}
Heorhiy Byelas and Alexandru~C. Telea.
\newblock Visualization of areas of interest in software architecture diagrams.
\newblock In {\em SoftVis '06: Proceedings of the 2006 ACM symposium on
  Software visualization}, pages 105--114, New York, NY, USA, 2006. ACM.

\bibitem{Cabr07a}
Bruno Cabral and Paulo Marques.
\newblock Exception handling: A field study in {Java} and {.NET}.
\newblock In {\em Proceedings of European Conference on Object-Oriented
  Programming (ECOOP'07)}, volume 4609 of {\em LNCS}, pages 151--175. Springer
  Verlag, 2007.

\bibitem{Dieh07a}
Stephan Diehl.
\newblock {\em Software Visualization}.
\newblock Springer-Verlag, Berlin Heidelberg, 2007.

\bibitem{Duca06c}
St\'ephane Ducasse, Tudor G\^irba, and Adrian Kuhn.
\newblock Distribution map.
\newblock In {\em Proceedings of 22nd IEEE International Conference on Software
  Maintenance (ICSM '06)}, pages 203--212, Los Alamitos CA, 2006. IEEE Computer
  Society.

\bibitem{Gree05d}
Orla Greevy, Michele Lanza, and Christoph Wysseier.
\newblock Visualizing feature interaction in {3-D}.
\newblock In {\em Proceedings of {VISSOFT} 2005 (3th IEEE International
  Workshop on Visualizing Software for Understanding)}, pages 114--119,
  September 2005.

\bibitem{Hite99a}
Kenneth Hite, Craig Neumeier, and Michael~S. Schiffer.
\newblock {\em GURPS Alternate Earths}, volume~2.
\newblock Steve Jackson Games, Austin, Texas, 1999.

\bibitem{Juck06a}
Susanne Jucknath-John and Dennis Graf.
\newblock Icon graphs: visualizing the evolution of large class models.
\newblock In {\em SoftVis '06: Proceedings of the 2006 ACM symposium on
  Software visualization}, pages 167--168, New York, NY, USA, 2006. ACM.

\bibitem{Kauf01b}
Michael Kaufmann and Dorothea Wagner.
\newblock {\em Drawing Graphs}.
\newblock Springer-Verlag, Berlin Heidelberg, 2001.

\bibitem{Kien07a}
Holger~M. Kienle and Hausi~A. Muller.
\newblock Requirements of software visualization tools: A literature survey.
\newblock {\em VISSOFT 2007. 4th IEEE International Workshop on Visualizing
  Software for Understanding and Analysis}, pages 2--9, 2007.

\bibitem{Kuhn07a}
Adrian Kuhn, St\'ephane Ducasse, and Tudor G\^irba.
\newblock Semantic clustering: Identifying topics in source code.
\newblock {\em Information and Software Technology}, 49(3):230--243, March
  2007.

\bibitem{Lang05a}
Guillaume Langelier, Houari Sahraoui, and Pierre Poulin.
\newblock Visualization-based analysis of quality for large-scale software
  systems.
\newblock In {\em ASE '05: Proceedings of the 20th IEEE/ACM international
  Conference on Automated software engineering}, pages 214--223, New York, NY,
  USA, 2005. ACM.

\bibitem{Lanz03d}
Michele Lanza and St\'ephane Ducasse.
\newblock Polymetric views---a lightweight visual approach to reverse
  engineering.
\newblock {\em Transactions on Software Engineering (TSE)}, 29(9):782--795,
  September 2003.

\bibitem{Marc01a}
Andrian Marcus and Jonathan~I. Maletic.
\newblock Identification of high-level concept clones in source code.
\newblock In {\em Proceedings of the 16th International Conference on Automated
  Software Engineering (ASE 2001)}, pages 107--114, November 2001.

\bibitem{Marc08a}
Andrian Marcus, Denys Poshyvanyk, and Rudolf Ferenc.
\newblock Using the conceptual cohesion of classes for fault prediction in
  object-oriented systems.
\newblock {\em IEEE Transactions on Software Engineering}, 34(2):287--300,
  2008.

\bibitem{Mesn05b}
C{\'e}dric Mesnage and Michele Lanza.
\newblock {White Coats}: Web-visualization of evolving software in {3D}.
\newblock {\em VISSOFT 2005. 3rd IEEE International Workshop on Visualizing
  Software for Understanding and Analysis}, 0:40--45, 2005.

\bibitem{Noac05a}
Andreas Noack and Claus Lewerentz.
\newblock A space of layout styles for hierarchical graph models of software
  systems.
\newblock In {\em SoftVis '05: Proceedings of the 2005 ACM symposium on
  Software visualization}, pages 155--164, New York, NY, USA, 2005. ACM.

\bibitem{Reis05a}
Steven~P. Reiss.
\newblock The paradox of software visualization.
\newblock {\em VISSOFT 2005. 3rd IEEE International Workshop on Visualizing
  Software for Understanding and Analysis}, page~19, 2005.

\bibitem{Sloc05a}
Terry~A. Slocum, Robert~B. McMaster, Fritz~C. Kessler, and Hugh~H. Howard.
\newblock {\em Thematic Carthography and Geographic Visualization}.
\newblock Pearson Prentice Hall, Upper Saddle River, New Jersey, 2005.

\bibitem{Stor05a}
Margaret-Anne~D. Storey, Davor \v{C}ubrani\'c, and Daniel~M. German.
\newblock On the use of visualization to support awareness of human activities
  in software development: a survey and a framework.
\newblock In {\em SoftVis'05: Proceedings of the 2005 ACM symposium on software
  visualization}, pages 193--202. ACM Press, 2005.

\bibitem{Stri05a}
Marc Strickert, Stefan Teichmann, Nese Sreenivasulu, and Udo Seiffert.
\newblock High-throughput multi-dimensional scaling {(HiT-MDS)} for
  {cDNA-Array} expression data.
\newblock In Wlodzislaw Duch, Janusz Kacprzyk, Erkki Oja, and Slawomir
  Zadrozny, editors, {\em ICANN}, volume 3696 of {\em Lecture Notes in Computer
  Science}, pages 625--633. Springer, 2005.

\bibitem{Term05a}
Maurice Termeer, Christian~F.J. Lange, Alexandru Telea, and Michel~R.V.
  Chaudron.
\newblock Visual exploration of combined architectural and metric information.
\newblock {\em VISSOFT 2005. 3rd IEEE International Workshop on Volume}, 0:11,
  2005.

\bibitem{Gude06a}
J\"urgen~Wolff v.~Gudenberg, A.~Niederle, M.~Ebner, and Holger Eichelberger.
\newblock Evolutionary layout of uml class diagrams.
\newblock In {\em SoftVis '06: Proceedings of the 2006 ACM symposium on
  Software visualization}, pages 163--164, New York, NY, USA, 2006. ACM.

\bibitem{Vasa07b}
Rajesh Vasa, Jean-Guy Schneider, and Oscar Nierstrasz.
\newblock The inevitable stability of software change.
\newblock In {\em Proceedings of 23rd IEEE International Conference on Software
  Maintenance (ICSM '07)}, pages 4--13, Los Alamitos CA, 2007. IEEE Computer
  Society.

\bibitem{Ware04a}
Colin Ware.
\newblock {\em Information Visualisation}.
\newblock Elsevier, Sansome Street, San Fransico, 2004.

\bibitem{Wett07b}
Richard Wettel and Michele Lanza.
\newblock Visualizing software systems as cities.
\newblock In {\em Proceedings of VISSOFT 2007 (4th IEEE International Workshop
  on Visualizing Software For Understanding and Analysis)}, pages 92--99, 2007.

\bibitem{Wise99a}
James~A. Wise.
\newblock The ecological approach to text visualization.
\newblock {\em J. Am. Soc. Inf. Sci.}, 50(13):1224--1233, 1999.

\end{thebibliography}
\end{document}